\documentclass[12pt]{article}

\newcommand{\beq}{\begin{equation}}
\newcommand{\eeq}{\end{equation}}
\newcommand{\beqa}{\begin{eqnarray}}
\newcommand{\eeqa}{\end{eqnarray}}
\newcommand{\beqar}{\begin{eqnarray*}}
\newcommand{\eeqar}{\end{eqnarray*}}

\def \ra {\rangle}

\begin{document}

\title{    \Large \bf
         ``Weighing'' a Closed System \\
and the Time-energy Uncertainty Principle
                 }
  \author{ Yakir Aharonov$^{(a,b)}$,
and Benni Reznik$^{(b)}$
\footnote{E-mail: reznik@post.tau.ac.il}
{\ } \\
(a) {\it  \small School of Physics and Astronomy, Tel Aviv
University, Tel Aviv 69978, Israel}\\
(b) {\it \small Department of Physics,
University of South Carolina, Columbia, SC 29208.}
}

\maketitle

\begin{abstract}
A gedanken-experiment is proposed for `weighing'' the total mass
of a closed system from within the system. We prove that for an
internal observer the time $\tau$, required to measure the total
energy with accuracy $\Delta E$, is bounded according to $\tau
\Delta E >\hbar $. This time-energy uncertainty principle for a
closed system follows from the measurement back-reaction on the
system.  We generally examine what other conserved observables are
in principle measurable  within a closed system and what are the
corresponding uncertainty relations.
\end {abstract}

\newpage

%The meaning of the time-energy uncertainty relation was an issue
%of debate in the early years of quantum theory \cite{abtime,peres}.
%The source of confusion may be traced to the special role
%played by time in quantum theory. Time is not represented
%by a time operator, it is a classical parameter.
%However time can be measured indirectly by observing a
%dynamical variable that is  linear in the time $t$,
%and is defined to be the clock time.

Time and frequency are classically two conjugate variables.
Nevertheless,  the interpretation of
the consequent quantum time-energy uncertainty relation is not
straightforward as for the case of other conjugate variables.
Aharonov and Bohm have shown that within quantum theory there is
no fundamental restriction on the  minimal time needed to measure the
total energy with given accuracy \cite{abtime}.
If the Hamiltonian of the system is known,
one can in principle setup a measurement of the Hamiltonian,
with arbitrary accuracy, at any time short as we please.
Instead, $\Delta\tau$ in the time-energy relation
\beq
\Delta \tau\Delta E\ge \hbar 
\eeq
can be interpreted as
the uncertainty
{\it caused to the internal time}  $\tau$ of the system due
to the measurement.

The Bohr-Einstein weighing  gedanken-experiment  \cite{jammer}
 illustrates  this interpretation.
The total mass of a closed box  (before and after the emission of the
photon) is there measured by weighing
the system in an external gravitational field.
The energy of the box  is then deduced from the
equivalence of mass and energy.
Bohr has shown that the process of weighing introduces a quantum uncertainty
in the location of the box  in the external gravitational field.
The uncertainty in the gravitational potential
leads in turn to an uncertainty in the internal time $\tau$ of the clock
within the box relative to the external time t \cite{unruh-opat}.

The purpose of this note is to offer another interpretation
of the time-uncertainty relation.
As long as the energy is measured with respect to a clock external to
the system, there is no fundamental restriction on the duration
of the measurement.
Suppose that an observer within a closed system measures the
total energy. We will argue that:
\begin{itemize}
\item
{\it The internal time,  $\tau$, needed to measure the total energy of an
isolated system, within a precision $\Delta E$, from within the system,
satisfies}  $\tau \Delta E \ge \hbar$.
\end{itemize}
Here $\tau$ is interpreted as the time shown by a physical clock within the
system, and $E$ is the total energy of the system {\it including the
internal clock}.

To illustrate this we first consider a gedanken-experiment
for measuring the total energy of an isolated system, by employing
gravity as in the Bohr-Einstein weighing experiment.
Let the system be a spherical shell of radius $R$, and
mass $M$, with an internal clock dynamical variable $\tau$.
At a certain clock time,
a test particle of mass $m\ll M$, which for simplicity we
take to be a spherical shell as well,
is ejected outwards with an initial velocity $v_0$
and  after traversing a distance $z_{max} \ll R$ is observed to fall
back to the shell surface at time $\tau$.
Classically, the mass of the shell can then be deduced from
$ M = {2R^2 v_0 /G\tau}$, where $G$ is Newton's constant.

However, the equivalence of energy and weight implies that
the clock rate must be affected by the test-shell according to
\beq
\tau(z)= t\biggl(1 + {\phi(z)\over c^2}  \biggr) 
\eeq
Here $\phi(z)$ is the gravitational potential at the position of the
clock $r=R$,
and $c$ is the velocity of light.
Note that $\phi(z)$ is a function of the hight, $z=r_{shell}-R$,
of the test shell.
Particularly, for  $z\ll R$,
the change in the potential at $r=R$ when the a shell location is $z$,
is given by
\beq
\delta \phi(z) = \phi(z)- \phi(z=0)= {Gm\over R^2}z  
\eeq
If the radial location of the shell has a quantum uncertainty
$\Delta z$, the above relation implies a
quantum uncertainty $\Delta \tau$ in the clock time. For weak gravitational fields,
 ${\phi\over c^2}\ll1$, and
\beq
{\Delta \tau \over \tau} = {Gm\over R^2}{\Delta z \over c^2} 
\eeq

The uncertainty  $\Delta z$ in the location of the test-shell
cannot be too small, because then the uncertainty of the
radial momentum of the shell becomes large.
If we like to measure the mass with an accuracy
$\Delta M$, the change in the impulse, $\delta p= \int Fd\tau\approx F\tau$,
caused by $\Delta M$ during the time $\tau$ must be
larger than the quantum  uncertainty in  the momentum of the test shell
\beq
{Gm \Delta M\over R^2} \tau > \Delta p_z 
\eeq
Combining the last two equations we obtain
\beq
\Delta \tau \Delta M > {1\over c^2} \Delta z\Delta p_z > \hbar .
\eeq
Finally, using the relation $\Delta E = \Delta M c^2$,
and the requirement $\tau >\Delta \tau$, we arrive to
\beq
\tau \Delta E > \hbar
\eeq

The time-energy uncertainty relation derived above follows from
the gravitational time dilation caused to the clock.
We will now show that this conclusion follows most generally,
irrespective of the details of the mechanism used,
whenever the total energy {\it including the internal clock energy},
is measured with respect to the internal clock time.

Let us consider an isolated ``box'' described by a Hamiltonian
$H_c+H_{box}$, where $H_c$ describes a clock, and $H_{box}$ the
rest of the system in the box. To describe a measurement
we will  couple the total energy to a measuring device with coordinate
$z$ and conjugate momentum $p$. For simplicity,
we can take the Hamiltonian of the measuring device as $H_{MD}=0$.
The total Hamiltonian including the von-Neumann measurement interaction is
\beq
H=  H_c+ H_{box} + {1\over 2}\biggl( g(\tau) H_c +H_c g(\tau)
+ 2 g(\tau)H_{box}      \biggr) z
\eeq
$g(\tau)$, is  the coupling function that is nonzero during the
measurement and is normalized: $\int g(\tau)d\tau= 1$. Since $H_c = -i\hbar {\partial \over \partial \tau}$
an appropriate ordering was assumed to keep the Hamiltonian Hermitian.

Suppose that the system is in an energy eigenstate,
\beq
H\Psi = E_0\Psi
\eeq
With the substitution
\beq
\Psi= \psi(\tau)u_E|z\ra
\eeq
where $H_{box}u_E=Eu_E$ and $|z\ra$ is an eigenstate of $z$, we get
\beq
{\partial \psi\over \partial \tau} =\biggl[ -{1\over 2} 
{z{dg\over d\tau}\over 1+ zg}
-i{E\tau\over \hbar }+ i {E_0/\hbar \over 1 + zg(\tau)} \biggr]\psi
\eeq
 and
\beq
\psi(\tau) = {1\over \sqrt{1+ g(\tau)z}}
e^{-i{E\tau\over \hbar}} e^{i{E_0\over \hbar} \int^\tau {d\tau'\over 1+g(\tau')z}}
\eeq
It can now be shown that only if
\beq
g(\tau)z \ll 1
\eeq
is satisfied, the solution $\psi(\tau)$ describes a measurement.
In this particular case
\beq
\Psi \simeq e^{-i{(E-E_0)\tau\over \hbar}} e^{-i{E_0z\over\hbar}
 \int g(\tau') d\tau'} u_E|z\ra
\eeq
Indeed the  last term, $\exp(-izE_0\int g(\tau') d\tau')$,
shifts the measuring device momentum $p$ by
\beq
\delta p = E_0 \int g(\tau')d\tau' = E_0
\eeq
If the duration of the measurement is $\tau_0$
the magnitude of the coupling function is  $g(\tau) \sim 1/ \tau_0$.
Since the accuracy $\Delta E_0$ of the
measurement is related to $z$ by $\Delta E_0= \Delta p \ge  \hbar /\Delta z
\sim \hbar /z$ we finally obtain that eq. (13) implies
\beq
\tau_0 \Delta E_0  \gg \hbar
\eeq
Therefore the measurement succeeds only if the duration $\tau$ of the
coupling satisfies the above uncertainty relation.

In passing let us compare the gravitational weighing
experiment and the von-Neumann measurement discussed above. In
both cases the measurement affects the rate of the clock.  In the
latter case, during the measurement the effective clock
Hamiltonian changes $H_c \rightarrow H_c(1+ z g)$. Therefore,  the
clock rate changes according to $\tau = t(1+ g z)$, and $g z$
plays here the role of the gravitational potential ${\phi(z)\over
c^2}$. The uncertainty of the clock caused by the test shell is
here due to the uncertainty of the coordinate $z$ conjugate to
the measuring device ``pointer'' $p$. In both cases the
uncertainty relation is due to the {\it measurement back-reaction}
on the clock. However a distinctive feature in the von-Neumann
measurement is that for a too small duration,  $\tau <{\hbar\over
\Delta E}$, the interaction does not yield the proper correlations
with the measuring device, i.e., the von-Neumann measurement
procedure fails \cite{casher-reznik}.

Finally, a more general perspective is provided by considering
 the general question of the observability of conserved
quantities from within a closed system. The weighing measurements
discussed here and the consequent time-energy uncertainty relation
are one special important case. However what is the general class
of conserved observables,  and what are the respective uncertainty
relations? We suggest that {\it every  scalar quantity within a
closed system is in principle measurable}, and generally gives
rise to analogous uncertainty relations.

Consider first a closed non-relativistic system. The symmetry
generators of Galilean boosts and rotations, are $G$ and $L$, and
of space and time translations are $P$ and $H$.  All four
generators are constants of motion, however they are not all
measurable within a closed system. As is well-known, observables
as position, velocity, angular momentum, etc, both in classical
mechanics as well as in quantum mechanics, are {\it relative}
observables. Indeed, we never measure the absolute position of a
particle, but the distance in between the particle and some other
object. Similarly, we never measure the angular momentum of a
particle along an absolute axis, but along a direction defined by
some other physical objects. Therefore the angular momentum of a
closed system can be measured only with respect to a point within
the system, say the location of the center of mass, and along a
direction defined by constitutes of the system. With respect to
the center of mass of a closed system
\beqa L= L_{cm} + L_{i}
\\ H= H_{cm} + H_{i}
\\ P=P_{cm} + P_i
\eeqa Since $L_i$ (along a certain direction) and $H_i$ are
scalars and since they are defined exclusively in terms of
internal variables they are internally measurable. By definition
$P_i$ must identically vanish.

Let us consider in more details the analogous  uncertainty
relation in a non-relativistic measurement of $L_i$. For
simplicity let our system be a rotating rigid disc of mass $M$.
The  axis of rotation can be located as the axis on which the
centrifugal forces vanish. Since distances are measured relative
to this axis, the moment of inertia,
 $(I=\sum m_ir_i^2)$, can also be measured.
Therefore, by measuring the angular velocity $\omega$ one can
deduce what is the angular momentum from $L_i=I\omega$. To this
end we will consider a measurement of the centrifugal force on a
test particle of mass $m\ll M$. We let $m$ slide along a radial
track with $\theta = constant$, with respect to the disc, and
measure the acceleration $a=\omega^2 r$. Classically this enables
us to determine the angular momentum.

For a quantum test particle, we note however that a quantum
uncertainty in its radial position $r$ introduces an uncertainty
in the contribution of the test particle to the total moment of
inertia $\Delta I = 2 mr\Delta r$. This in turn causes, via the
conservation of angular momentum, an uncertainty $\Delta \omega
\sim {I\over \omega}\Delta I$ in the angular momentum. Hence after
time $T$ the relative angle of the disc becomes uncertain with
respect to an external frame of reference by the amount
\beq
\Delta \theta = T \Delta \omega
\eeq

On the other hand, we cannot have very small $\Delta r$ because
then the uncertainty in the radial momentum $\Delta p$ becomes
large. Indeed, we must also require that the change in the
impulse, $\int F dt \simeq m\omega^2 r T$, when $\omega$ is
measured with precision $\Delta \omega$, must be larger than the
uncertainty in the radial momentum of the particle
\beq
 2 m\omega rT\Delta \omega  >  \Delta p \eeq
Combining the last two equations and using $\Delta L \simeq
I\Delta\omega \simeq \omega \Delta I$, we finally obtain

\beq \Delta \theta \Delta L > \hbar \eeq

Hence a  measurement of $L$ with accuracy $\Delta L$ causes a
minimal uncertainty $\Delta \theta> \hbar/\Delta L$ in the
relative angle of the disc and an external frame. That is
 in complete analogy with our previous discussion; there, weighing
 the system has caused an  uncertainty in the internal
 time.

%By sending a test particle from the center of mass outwards and
%measuring the angular shift $\theta - theta_0$ during  the time $\tau-\tau_0$
%we finds $\omega = {\theta-theta_0\over \tau-\tau_0}$.
%However a  measurement of  $\theta$ with accuracy $\Delta \theta$
%causes an uncertainty,  $\Delta L_{test-particle} \ge \hbar\Delta
%\theta$, in the angular momentum of the test particle.
%Since $L_i$ is conserved, the accuracy of the measurement is
%at best $\Delta L_i \ge   \Delta L_{test-particle}$.
%Finally, since we must have also $\theta-\theta_0 \gg \Delta\theta$
%we get
%\beq
%(\theta - \theta_0) \Delta L_i > \hbar.
%\eeq
%Hence a  measurement of $L_i$ with accuracy $\Delta L_i$ requires
%a minimal rotation $\theta-\theta_0$.

In a relativistic theory the 10 generators of boosts, rotations,
and space-time translations, form the Poincar\'e group. The
observables in a closed system must be scalars with respect to
Poincar\'e group. It is well known that the group has two Casimir
invariants $C_1= P_\mu P^\mu =m^2$ where $P_\mu$ is the
energy-momentum four-vector, and $C_2 = W_\mu W^\mu=-m^2s(s+1)$
where $W_\mu$ is the Pauli-Lubanski pseudo-vector. The mass and
spin are two scalars.  Hence in a relativistic system, the
non-relativistic internal energy $H_i$ becomes the rest mass
$m=\sqrt{E^2-p^2}$, and the internal angular momentum corresponds
to the spin $s$. Similarly in our weighing experiment the total
energy is measured with respect to the rest mass of the shell
system, hence what we have measured is the rest mass of a closed
system.

In conclusion we have shown that the energy of a closed system can
be measured from within the system. However while quantum theory
poses no limitation on the duration of the measurement of energy
in an open system, from within a closed system the duration of the
measurement satisfies a time-energy uncertainty. Similar
uncertainty relations can be found for other conserved
observables.

 {\bf Acknowledgment}
We acknowledge the support of  the Basic
Research Foundation,  grant 614/95,
administered by the Israel Academy of Sciences and Humanities.
The work of Y. A. was supported by NSF grant PHY-9601280.

\vfill \eject


\begin{thebibliography}{99}
\bibitem{abtime}
Y. Aharonov and D. Bohm, Phys. Rev. {\bf 122}, 1649 (1961)

\bibitem{jammer}
For a review of the Bohr-Einstein debate and for further references see
M. Jammer, {\it The Philosophy of Quantum Mechanics} Wiley, New York 1974,
see also {\it Quantum Theory of Measurement}, ed. J. A. Wheeler and W. H.
Zurek, Princeton Univ. Press, Princeton (1983).


\bibitem{unruh-opat}
This is a direct consequence of the possibility that energy has weight and
is independent of the details
of the theory of general relativity:
W. G. Unruh, and G. I. Opat, Am. J. Phys. {\bf47}, 743 (1979).

\bibitem{casher-reznik}
For further discussion on the role of the measurement
back-reaction on clocks in a closed systems:
A. Casher and B. Reznik, in preparation.


\end{thebibliography}
\end{document}